\documentclass[aps,nofootinbib,prd,preprintnumbers]{revtex4}
\usepackage{amstext}
\usepackage{amssymb}
\usepackage{graphicx}
\usepackage{rotating}
\usepackage{gensymb}
\usepackage[format=plain,singlelinecheck = false, format= hang, justification=raggedright]{caption}

\usepackage{multirow} 
\usepackage[T1]{fontenc} 
\usepackage{slashed}
\usepackage{caption}
\usepackage{float}
\usepackage[normalem]{ulem}

\usepackage[usenames,dvipsnames]{color}
\definecolor{mightnightblue}{RGB}{25,25,112}
\usepackage{hyperref}
 \hypersetup{
     colorlinks=true,
     linkcolor= Black,
     citecolor=mightnightblue,
    urlcolor=mightnightblue
    }

\newcommand {\black} {\color{black}}

\def\gsim{\raise0.3ex\hbox{$\;>$\kern-0.75em\raise-1.1ex\hbox{$\sim\;$}}}
\def\lsim{\raise0.3ex\hbox{$\;<$\kern-0.75em\raise-1.1ex\hbox{$\sim\;$}}}
\newcommand {\ignore}[1]{}

\newcommand{\eVq}  {\text{eV}^2}

\newcommand{\AddrAHEP}{%
  AHEP Group, Institut de F\'{i}sica Corpuscular --
  C.S.I.C./Universitat de Val\`{e}ncia, Parc Cientific de Paterna.\\
 C/ Catedratico Jos\'e Beltr\'an, 2 E-46980 Paterna (Val\`{e}ncia) - SPAIN}

   \newcommand{\unicamp}{Instituto de F\'{i}sica Gleb Wataghin - UNICAMP,
13083-859, Campinas, SP, Brazil}
  \newcommand{\virginia}{Center for Neutrino Physics, Virginia Tech,
  Blacksburg, VA 24061, USA}

\begin{document}


\title{Status of neutrino oscillations 2018: \\
 first hint for normal mass ordering and improved CP sensitivity~\footnote{https://globalfit.astroparticles.es/}}

\author{P. F. de Salas$^1$}
\author{D. V. Forero~$^{2,3}$}\email{dvanegas@ifi.unicamp.br}
\author{C. A. Ternes$^1$}
\author{M. T{\'o}rtola~$^1$}\email{mariam@ific.uv.es}  
\author{J.~W.~F.~Valle~$^1$} \email{valle@ific.uv.es} 
\affiliation{$^1$~\AddrAHEP}
\affiliation{$^2$~\unicamp}
\affiliation{$^3$~\virginia}


\vskip 2cm

\begin{abstract}

  We present a new global fit of neutrino oscillation parameters
  within the simplest three-neutrino picture, including new data which
  appeared since our previous analysis~\cite{Forero:2014bxa}. In this
  update we include new long--baseline neutrino data involving the
  antineutrino channel in T2K, as well as new data in the neutrino channel, data from NO$\nu$A, as well
  as new reactor data, such as the Daya Bay 1230 days electron
  antineutrino disappearance spectrum data and the 1500 live days
  prompt spectrum from RENO, as well as new Double Chooz data.
  We also include atmospheric neutrino data from the IceCube DeepCore
  and ANTARES neutrino telescopes and from Super-Kamiokande.
  Finally, we also update our solar oscillation analysis by including
  the 2055-day day/night spectrum from the fourth phase of the
  Super-Kamiokande experiment.
With the new data we find a preference for the atmospheric angle in
the upper octant for both neutrino mass orderings, with maximal mixing
allowed at $\Delta\chi^2 = 1.6 \, (3.2)$ for normal (inverted) ordering.
We also obtain a strong preference for values of the CP phase $\delta$ in the
range $[\pi,2\pi]$, excluding values close to $\pi/2$ at
more than 4$\sigma$.
More remarkably, our global analysis shows for the first time hints in
favour of the normal mass ordering over the inverted one at more than
3$\sigma$.
We discuss in detail the origin of the mass ordering, CP violation and
octant sensitivities, analyzing the interplay among the different
neutrino data samples.
\end{abstract}
\maketitle

\black
\section{Introduction}

The discovery of neutrino oscillations constitutes a major milestone
in astro and particle physics over the last few decades. 
Solar and atmospheric neutrino studies were the first to give a 
convincing evidence for neutrino
conversion~\cite{McDonald:2016ixn,Kajita:2016cak}.
By studying the distortion in the neutrino spectra, laboratory
experiments based at reactors and accelerators have played a key role
in selecting neutrino oscillations as the conversion mechanism at
work.
Reactor and accelerator experiments have now brought the field of
neutrino oscillations to the precision era, contributing significantly
to sharpen the determination of the oscillation
parameters~\cite{Maltoni:2002aw,Maltoni:2004ei,Nunokawa:2007qh,GonzalezGarcia:2012sz,Fogli:2012ua,Balantekin:2013tqa}.
Particularly relevant was the input of the KamLAND experiment in
elucidating the nature of the solution to the solar neutrino
puzzle~\cite{eguchi:2002dm,Abe:2008aa}. Indeed, KamLAND measurements
have ruled out alternative mechanisms involving spin flavor
precession~\cite{Miranda:2003yh,miranda:2004nz} as well as 
  nonstandard neutrino interaction (NSI) solutions to the solar
  neutrino problem~\cite{guzzo:2001mi}. Such NSI-only
  scenarios as well as all other more exotic hypotheses are all ruled
  out by KamLAND~\cite{pakvasa:2003zv,Maltoni:2004ei}.

Precision tests of the oscillation picture have already a long
history, and remain as timely as ever.
Indeed, one can probe neutrino NSI with
atmospheric~\cite{fornengo:2001pm} as well as solar neutrino data~\cite{bolanos:2008km,palazzo:2009rb},
where the robustness of the solar neutrino oscillation description has
been questioned~\cite{Miranda:2004nb,Escrihuela:2009up}.
There have been a variety of studies scrutinizing the possible role of
NSI in various neutrino oscillation
setups~\cite{Huber:2001de,Huber:2001zw,Huber:2002bi,friedland:2004ah,barranco:2005ps,Bandyopadhyay:2007kx,estebanpretel:2008qi,Escrihuela:2011cf,Agarwalla:2014bsa,deGouvea:2015ndi,Coloma:2015kiu,Farzan:2016wym,Forero:2016cmb,deSalas:2016svi,Coloma:2017egw}.
Likewise, although already constrained by experiment, the effect of
neutrino non-unitarity of the lepton mixing matrix, expected if
neutrino masses arise \textit{a la
  seesaw}~\cite{valle:1987gv,Escrihuela:2015wra,Miranda:2016ptb},
could lead to important ambiguities in probing CP violation in
neutrino oscillations~\cite{Miranda:2016wdr}, as well as opportunities
for probing novel effects. These need to be taken up seriously in the
design of future oscillation
experiments~\cite{Ge:2016xya,Blennow:2016jkn,Escrihuela:2016ube}.
One example are neutrino factories, which also provide a potential
testing ground for the non-unitarity of the neutrino mixing
matrix~\cite{Goswami:2008mi,Antusch:2009pm}.
 
Similarly, neutrino magnetic moment interactions in turbulent
convective-zone magnetic fields would induce an enhanced solar
antineutrino flux, to which KamLAND observations are
sensitive~\cite{Miranda:2003yh,miranda:2004nz}.
Likewise, radiative-zone random magnetic fields~\cite{burgess:2003fj}
would induce sizeable density fluctuations, capable of affecting
neutrino propagation in a significant
manner~\cite{nunokawa:1996qu,balantekin:1996pp}.  However, under the
hypothesis of CPT conservation, KamLAND constrains the effect of
potentially large density fluctuations on solar neutrino
oscillations~\cite{burgess:2002we,burgess:2003su}.

Here we reconsider the determination of neutrino oscillation
parameters within the simplest three-neutrino picture, in the light of
new data that appeared since our previous published global
analysis~\cite{Forero:2014bxa}. These include new long--baseline
disappearance and appearance data involving the antineutrino channel
in T2K \cite{Abe:2017bay,Abe:2017uxa}, an updated dataset in the neutrino mode~\cite{t2k-hartz}, as well as
disappearance and appearance neutrino data from 
NO$\nu$A~\cite{Adamson:2017qqn,Adamson:2017gxd,nova-CERN}. 
 Turning to reactors, we have included the electron
  antineutrino disappearance spectrum of Daya Bay corresponding to
  1230 days of data~\cite{An:2016ses}, the 1500 live days prompt
  reactor spectra from RENO~\cite{Seo:2017ozk,Pac:2018scx} 
  as well as the Double Chooz event energy spectrum from the far-I and far-II data
  periods~\cite{dc-MORIOND:16}.
  Concerning atmospheric neutrinos, we have included data from the
  IceCube DeepCore~\cite{Aartsen:2014yll} and
  ANTARES~\cite{AdrianMartinez:2012ph} neutrino telescopes, properly
  taking into account the relevant matter effects in the neutrino
  propagation inside the Earth.
  Given the difficulties to analyze the most recent atmospheric
  neutrino data from Super-Kamiokande with the public information
  available, we directly include the $\chi^2$--tables provided by the
  Super-Kamiokande collaboration, corresponding to the combination of
  the four run periods of the experiment~\cite{Abe:2017aap}.
  Finally, we have also updated our solar oscillation analysis by
  including the 2055-day day/night spectrum from the fourth phase of
  the Super-Kamiokande experiment~\cite{Nakano:PhD}.  \black

\section{New Experiments}
\label{sec:new-experiments}

In this section we present a brief description of the NO$\nu$A
  long--baseline accelerator neutrino experiment as well as the
neutrino telescopes ANTARES and IceCube DeepCore which were
not included in the previous global fit~\cite{Forero:2014bxa}. 

\subsection*{The ANTARES neutrino telescope}
\label{sec:antar-neutr-telesc}

ANTARES is a deep sea neutrino telescope located at the Mediterranean
Sea, near Toulon (France). It consists of 12 lines with 75 optical
modules each, covering a height of 350m and anchored at the sea floor
at a depth of about $2.5\,\mathrm{km}$, with a separation of around
$70$~m between neighboring modules. 
The neutrino detection is based on the Cherenkov light emitted when
the charged leptons produced by the neutrino interactions move through
the water.
Although ANTARES was not designed to contribute to the determination
of the oscillation parameters, it was the first large volume
Cherenkov-based neutrino telescope performing such analysis with
atmospheric neutrinos. They managed to do it as a result of an
important reduction of their threshold energy, from 50 GeV, when only
multi-line events are considered, to 20 GeV for single-line events.

\subsection*{IceCube DeepCore}
\label{sec:icecube-deepcore}

IceCube is a $1\, \mathrm{km}^3$ multipurpose neutrino telescope
placed near the Amundsen-Scott South Pole Station, buried beneath the
surface and extending up to a depth of about 2500~meters.
Similarly to ANTARES, it uses Cherenkov light to detect high energy
neutrinos, with the difference that IceCube uses the polar ice as the
medium where this light is produced. It has 86 strings with 60 digital
optical modules (DOMs) each, placed at a depth that goes from
$1450\,\mathrm{m}$ to $2450\,\mathrm{m}$ into the ice.
In this analysis we use the data from DeepCore, a denser region of
strings inside IceCube, designed to measure the atmospheric neutrino
flux at low energies.  The observed energy lies between 6.3~GeV and
56.2~GeV, way below the energy threshold of IceCube, which is about
100~GeV.

\subsection*{The NO$\nu$A experiment}
\label{sec:nonua-experiment}

The NO$\nu$A experiment is a long--baseline neutrino oscillation
facility, with a 810 km baseline, which makes it the biggest long
baseline experiment to date. It was designed to observe
$\nu_\mu$-disappearance as well as $\nu_e$-appearance in both neutrino
and antineutrino channels.  In order to accomplish this, it uses an
intense and (nearly) pure beam of $\nu_\mu$ generated at the Fermilab
accelerator complex.
These neutrinos go through the Earth to northern Minnesota, 810 km
away, to be detected at the Ash River far detector.  The NO$\nu$A
experiment has collected an equivalent of $8.85\times 10^{20}$ protons
on target of data in the neutrino mode and is now taking data with the
antineutrino beam. Because of its 810 km baseline, it is more
sensitive to matter effects than the T2K experiment. 
With further data taking, this may translate into a better sensitivity
to the neutrino mass ordering.  The detectors are 14 mrad off-axis,
which results in a narrow neutrino energy spectrum, peaked around
2~GeV, which coincides with the oscillation maximum for
$\nu_\mu \to \nu_e$ oscillations.

\black
\section{New data}
\label{sec:new-data}

We now describe the new data samples used in this updated global
neutrino oscillation analysis.

\subsection*{Updated solar neutrino data sample}
\label{sec:updat-solar-neutr}

We have updated our solar oscillation analysis including the 2055-day
D/N (day/night) spectrum from the fourth phase of the Super-Kamiokande
experiment, according to Ref.~\cite{Nakano:PhD}. This new sample
includes the D/N energy spectrum above 3.5 MeV collected along 2055
days, from September 2008 to April 2015.  The signal observed
corresponds to a $^8$B solar neutrino flux of
$2.314 \pm 0.018 \mathrm{(stat)} \pm 0.039 \mathrm{(syst)}\times 10^6$
cm$^{-2}$s$^{-1}$.
  The measured D/N asymmetry during this period is determined as
  $A_{DN} = \left[-3.1\pm1.6 (\mathrm{stat}) \pm
    1.4(\mathrm{syst})\right]$
  \%, at 1.5$\sigma$ from zero.  Thanks to the increasing accuracy,
  this result combined with the observed D/N asymmetry in the three
  previous phases of Super-K, provides an indirect indication for
  matter-enhanced neutrino oscillations inside the Earth.
  Apart from small differences in the values of the oscillation
  parameters, the main results concerning the neutrino oscillation
  parameters remain intact with respect to our previous analysis in
  \cite{Forero:2014bxa}, in particular the fact that maximal solar
  neutrino mixing is highly disfavored{\footnote{The reanalysis
      of KamLAND data in the light of the recently observed "bump" in
      the reactor antineutrino spectrum might produce small deviations
      of the solar oscillation parameters, as obtained in Ref.
      \cite{Capozzi:2016rtj}}}.

\subsection*{New data from Daya Bay}
\label{sec:daya-bay}

Daya Bay is a multi-core and multi-detector experiment, with eight
$20\,\text{ton}$ Gd-doped liquid scintillator antineutrino detectors
(ADs) located at three experimental halls (EHs). At EH1 and EH2, two
ADs were deployed while the remaining ADs were assigned to the far
site, EH3. The thermal power of each reactor is
$2.9\,\text{GW}_{\text{th}}$ and the baseline to the near and far
sites (EH1 and EH2) are in the range $0.35-0.6$~km and $1.5-1.9$~km,
respectively. After $1230$~days of data taking, Daya Bay has measured
approximately two hundred thousand inverse beta decay events at the
far site. Thanks to the large statistics and the reduction of
systematical errors, due to having several functionally identical ADs,
Daya Bay has provided the most precise determination of the reactor
mixing angle to date.

In this analysis, we have included the antineutrino event energy
spectra from the three EHs.  Systematical errors accounting for total
and detector normalization, as well as core-related errors and energy
scale errors were included in the analysis.  Systematical errors
accounting for the background normalization in each experimental hall
have been also included in the analysis, where we have used the
background expectations from the ancillary files from
Ref.~\cite{An:2016ses}.

\subsection*{New data from RENO} 

The RENO experiment has recently reported $1500$~live days of data
from antineutrinos produced at six reactor cores each one with a
$\sim 2.8\,\text{GW}_{\text{th}}$ thermal power.  The experiment
detects neutrinos at a near and at a far detector (each detector with
16 ton of fiducial mass) located at $0.294$~km and $1.383$~km
from the line joining the six reactor cores, respectively\footnote{The
  exact detector to reactor distances from Ref.~\cite{Ahn:2010vy} were
  used in our simulation.}.
Thanks to the improved precision, the spectral fit analysis of RENO
data is now sensitive to the neutrino oscillation phase, as reported
in Refs.~\cite{Seo:2017ozk,Pac:2018scx}. 
In our analysis, we have considered the near and far detector event
energy distribution.  We have fitted the measured energy spectrum at
each detector after the subtraction of the background, normalizing our
simulation to the expected spectra reported by the RENO
collaboration. 
Systematical errors accounting for core-related ($0.9\%$ for each
core) and detector uncertainties ($0.2\%$ for each
detector)~\cite{RENO:2015ksa}, have been included in our analysis in
the form of nuisance parameters.  We have also included a nuisance
parameter accounting for the total normalization uncertainty, that has
been left completely free in the analysis.
\begin{figure}[t!]
\includegraphics[width=0.8\textwidth]{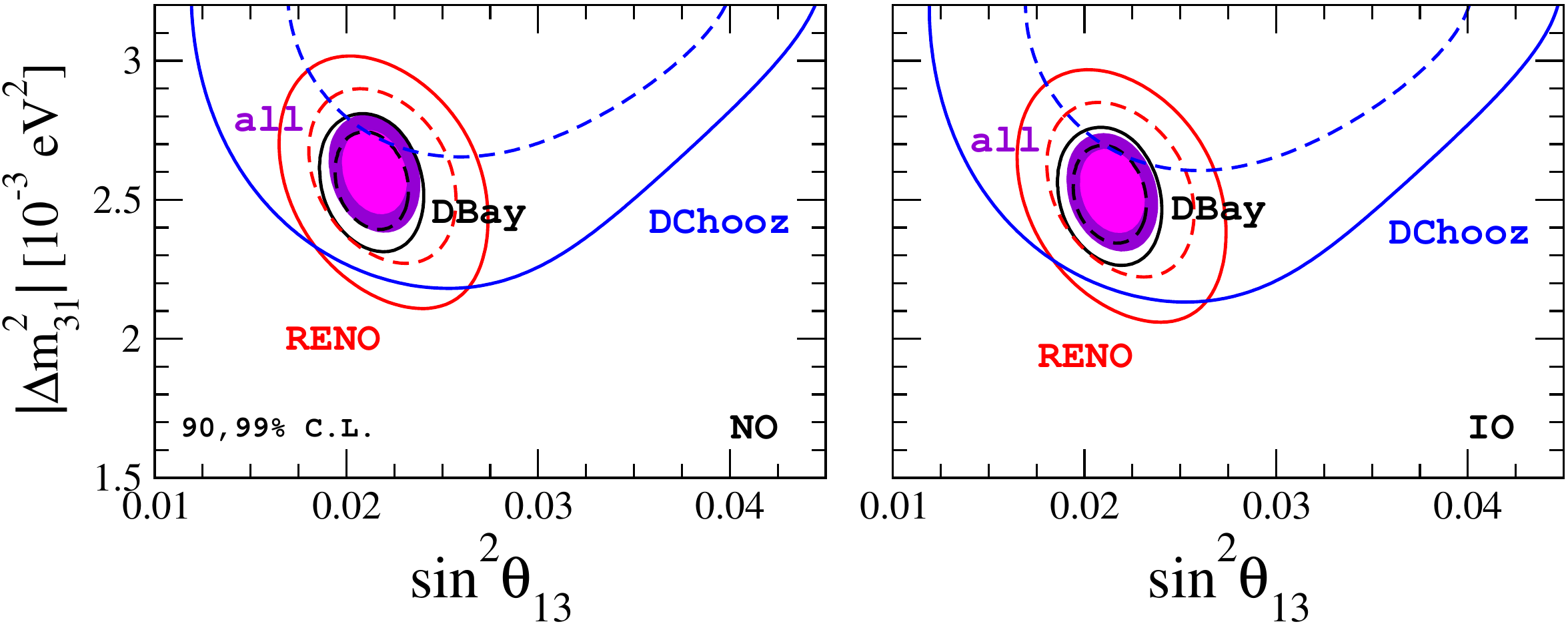}
\caption{90 and 99\% C.L. (2 d.o.f.) allowed regions at the $\sin^2\theta_{13}$--
  $\Delta m^2_{31}$ plane from individual reactor neutrino experiments
  (dashed and solid lines) and from the combination of the three
  experiments (coloured regions).  The  left (right) panels
  correspond to normal (inverted) mass ordering.}
\label{fig:reactor}
\end{figure}

\black
\subsection*{New data from Double Chooz} 
\label{sec:double-chooz}

The Double Chooz experiment detects antineutrinos produced at two reactor
cores with a $2\times 4.27\,\text{GW}_{\text{th}}$ total thermal power
with a near and far detector of 8 ton fiducial mass each, located 
at $0.4$~km and $1.05$~km, respectively. 
The data set considered in this analysis corresponds to $461$~days of
data with far detector only (far-I) plus $212$~days of far detector
data with a near detector (far-II), as reported in
Ref.~\cite{dc-MORIOND:16}\footnote{Even though more recent data has
  been presented at the Moriond conference~\cite{dc-MORIOND:17}, the collaboration is still trying to
  understand better their systematics. For this reason we have only
  included the previous data from~\cite{dc-MORIOND:16} in our
  analysis.} .
The event energy spectrum from the far-I
and far-II data periods were included in the analysis. 
Systematical errors considered in our simulation account for the signal
and background normalization as well as for the total normalization. The total background has been
extracted from the  data reported in Ref.~\cite{dc-MORIOND:16}.
The results of the analysis of the three reactor experiments are
given in Fig.\ref{fig:reactor} and will be discussed in detail in the
next section.
  
\subsection*{Atmospheric data from ANTARES}
\label{sec:antares}

We analyze atmospheric data from the ANTARES collaboration following
Ref.~\cite{AdrianMartinez:2012ph}, taking also into account matter
effects, and including electron neutrino and neutral current
interaction events.
In order to calibrate our simulation we have first reproduced very
well the analysis performed by the collaboration using their
assumptions and approximations.
Afterwards we have included neutral current interactions and matter
effects to our simulation. In Fig.~\ref{fig:all-atm} we plot the
allowed regions in the atmospheric parameters at 90 and 99\% C.L. from
our analysis of ANTARES data.
One sees the regions are still very large and therefore the
sensitivity is not competitive with the other experiments, described
in the following sections. It is expected that the
ANTARES collaboration will update their analysis with more data,
hopefully improving their sensitivity to the atmospheric neutrino
oscillation parameters.

\subsection*{Atmospheric data from IceCube DeepCore}
\label{sec:icecube-deepcore-1}

In order to determine the atmospheric neutrino oscillation parameters,
in this simulation we use data published by IceCube DeepCore in
Ref.~\cite{Aartsen:2014yll}, analyzed following all the updates
presented by the collaboration.
Neutrino data are presented in 64 bins, with 8 energy-bins and 8 bins
in zenith-angle, see \cite{IC-www}.
  Tables with systematic detector uncertainties, optical efficiencies
  and uncertainties produced through scattering at holes opened in the
  ice for the depletion of the DOMs are also provided.  The fluxes for
  atmospheric neutrinos are taken from \cite{Honda:2015fha,Honda-www}.
  We perform the numerical integration in matter using the Preliminary
  Reference Earth Model (PREM)~\cite{dziewonski:1981xy}. 
  In Fig.~{\ref{fig:all-atm} we compare the allowed regions in the
    atmospheric neutrino oscillation parameters $\sin^2\theta_{23}$
    and $\Delta m^2_{31}$ obtained from ANTARES, DeepCore and
    Super-Kamiokande phases I-IV at 90 and 99\% confidence
    level.
    As discussed in the next section, with such new analysis, DeepCore
    data are becoming competitive with those of long--baseline
    experiments NO$\nu$A or T2K in the determination of the
    atmospheric neutrino oscillation parameters \footnote{ The recent
      reanalysis of DeepCore data performed by the IceCube
      collaboration in Ref.~\cite{Aartsen:2017nmd} shows improved
      sensitivity to the atmospheric neutrino oscillation
      parameters. However, the details of this reanalysis are not yet
      publicly available, so this improvement cannot be incorporated
      in our global fit.}.

\begin{figure}
\centering
\includegraphics[width=0.8\textwidth]{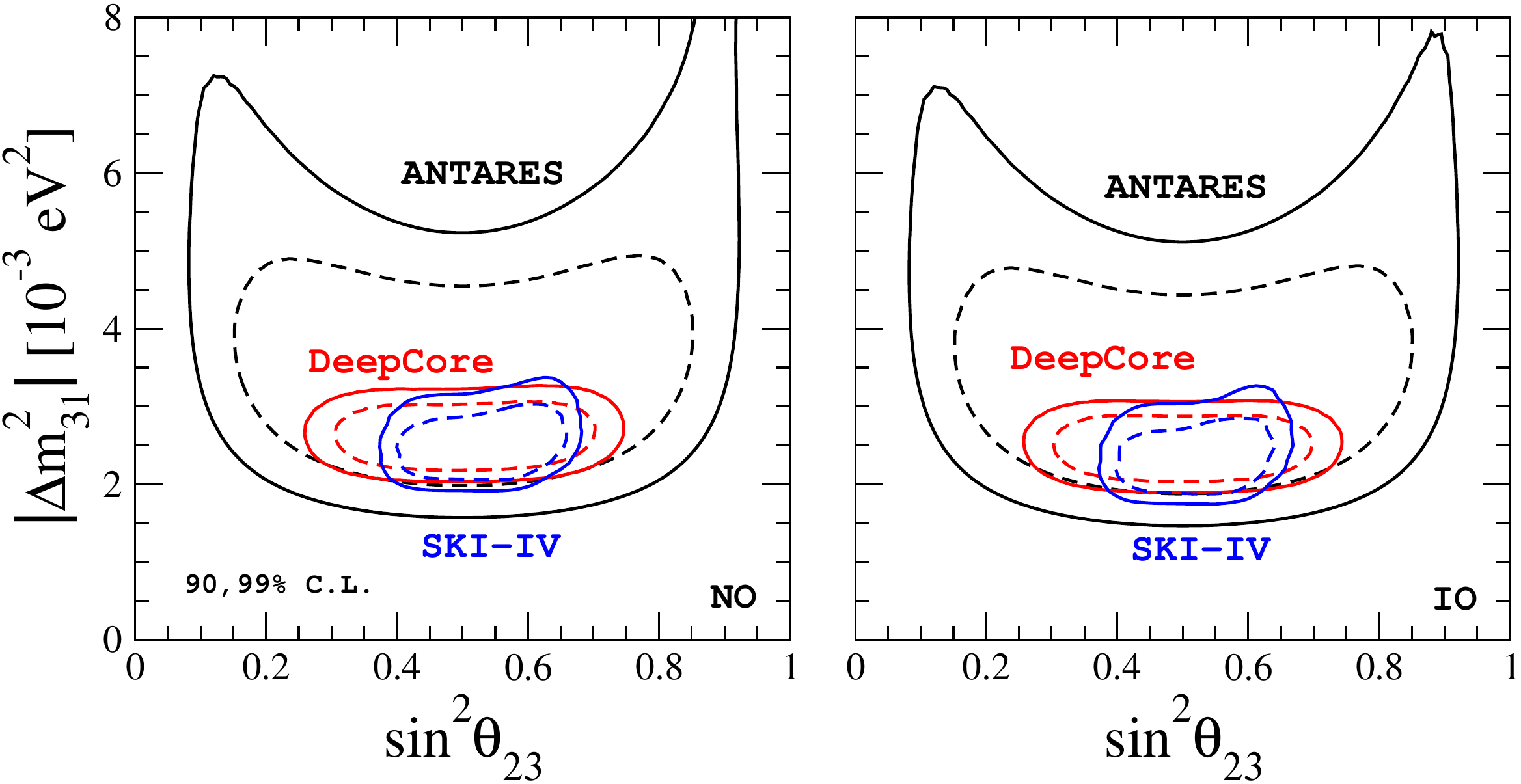}
\caption{
90 and 99\% C.L. (2 d.o.f.) allowed regions at the $\sin^2\theta_{23}$--
  $\Delta m^2_{31}$ plane obtained from the atmospheric neutrino
  experiments for normal  (left) and
  inverted ordering (right).}
\label{fig:all-atm}
\end{figure}

\subsection*{Atmospheric data from Super-Kamiokande}
\label{sec:antares}

In this work we include the most recent atmospheric neutrino results
from the Super-Kamiokande experiment~\cite{Abe:2017aap}, corresponding
to the combined analysis of phases I to IV of the experiment, with a
total of 328 kton-year exposure of the detector. 
The data analysis performed by the Super-K collaboration, optimized to
enhance the sensitivity to the neutrino mass ordering, 
  includes the impact of the atmospheric oscillation parameters as
  well as the reactor angle and the CP phase.
  As stressed in~\cite{Esteban:2016qun}, the most recent atmospheric
  neutrino Super-K data samples are not presented in a form that
  allows a reliable use outside the collaboration. Therefore, here we
  follow the same procedure adopted in previous publications
  (Refs.~\cite{Forero:2014bxa,Schwetz:2011qt,Tortola:2012te}) which
  consists of directly incorporating to our global neutrino analysis
  the $\chi^2$--tables provided by the Super-K
  collaboration~\cite{SKIV-tabs}, obtained in Ref.~\cite{Abe:2017aap}.

\subsection*{New long--baseline data from T2K}
\label{sec:new-T2K}

In addition to the data used in the neutrino oscillation global fit
published in Ref.~\cite{Forero:2014bxa}, the T2K collaboration has new
results in the neutrino mode. 
Therefore, this updated analysis includes the latest T2K antineutrino
sample as well as their updated neutrino data, as published in
Refs.~\cite{Abe:2017bay,Abe:2017uxa,t2k-hartz}.
With an accumulated statistics of 14.6$\times 10^{20}$ POT in the
neutrino run, the T2K collaboration now observes 240 disappearance and
74+15 appearance (charged current quasi-elastic and charged current
single-pion, respectively) neutrino events. Note, however, that the
CC-1$\pi$ appearance events have not been included in our simulation.
In the antineutrino channel, with 7.6$\times 10^{20}$ POT, a total of
68 disappearance $\bar{\nu}_\mu$ events and 7 appearance $\bar{\nu}_e$
events were recorded.  
In the present analysis we have included the newest neutrino fluxes
in Super-K provided by the T2K web page~\cite{T2K-flux}. 
The simulation of the experiment and the statistical analysis were
performed with the GLoBES package~\cite{Huber:2004ka,Huber:2007ji},
including all systematic uncertainties reported in
Ref.~\cite{t2k-hartz}.  

Notice that T2K has already achieved some CP sensitivity, as seen in
Fig.~\ref{fig:del-T2K-NOvA}.  Indeed, thanks to the combination of the
results in the neutrino and the antineutrino channel, T2K is the first
experiment able to exclude on its own certain values of the CP phase
 at more than $2\sigma$ for normal ordering (NO), and even at 3$\sigma$ for inverted 
ordering (IO). The allowed regions for other oscillation parameters, such
as $\theta_{13}$ and $\Delta m^2_{31}$, are found to be consistent
with the reactor experiments.

\subsection*{New long--baseline data from NO$\nu$A} 
\label{sec:first-data-from}

In our  global fit we also include the latest results for
$\nu_\mu$-disappearance and $\nu_e$-appearance of the NO$\nu$A
experiment.
NO$\nu$A has recently published the results of their neutrino run with
an accumulated statistics of 8.85$\times 10^{20}$
POT~\cite{nova-CERN}. 
In the disappearance channel, a total of 126 events have been observed,
while 763 events were expected under the no-oscillation hypothesis. In the
appearance channel, a total of 66 events have been detected.
The neutrino oscillation analysis reported by the NO$\nu$A
collaboration imposing a prior on $\theta_{13}$ slightly disfavors
inverted mass ordering, with a significance of approximately
2$\sigma$.
Our simulation of the NO$\nu$A experiment has been performed using
GLoBES\cite{Huber:2004ka,Huber:2007ji}, including all the systematic
errors reported in \cite{Adamson:2017qqn,Adamson:2017gxd} and updated
in Ref.~\cite{nova-CERN}.

\black

In Fig.~{\ref{fig:LBL} we compare the restrictions on
the atmospheric neutrino parameters derived from long--baseline
accelerator data coming from the T2K, NO$\nu$A and MINOS
experiments, at 90 and 99\% confidence level.
Further results are summarized in
Figs.~\ref{fig:del-T2K-NOvA},~\ref{fig:sq23.sq13.del},~\ref{fig:sq23.sq13},~\ref{fig:panel-dchi2}
and \ref{fig:2dim} and discussed in the following section.
\begin{figure}[t]
 \centering
   \includegraphics[width=0.8\textwidth]{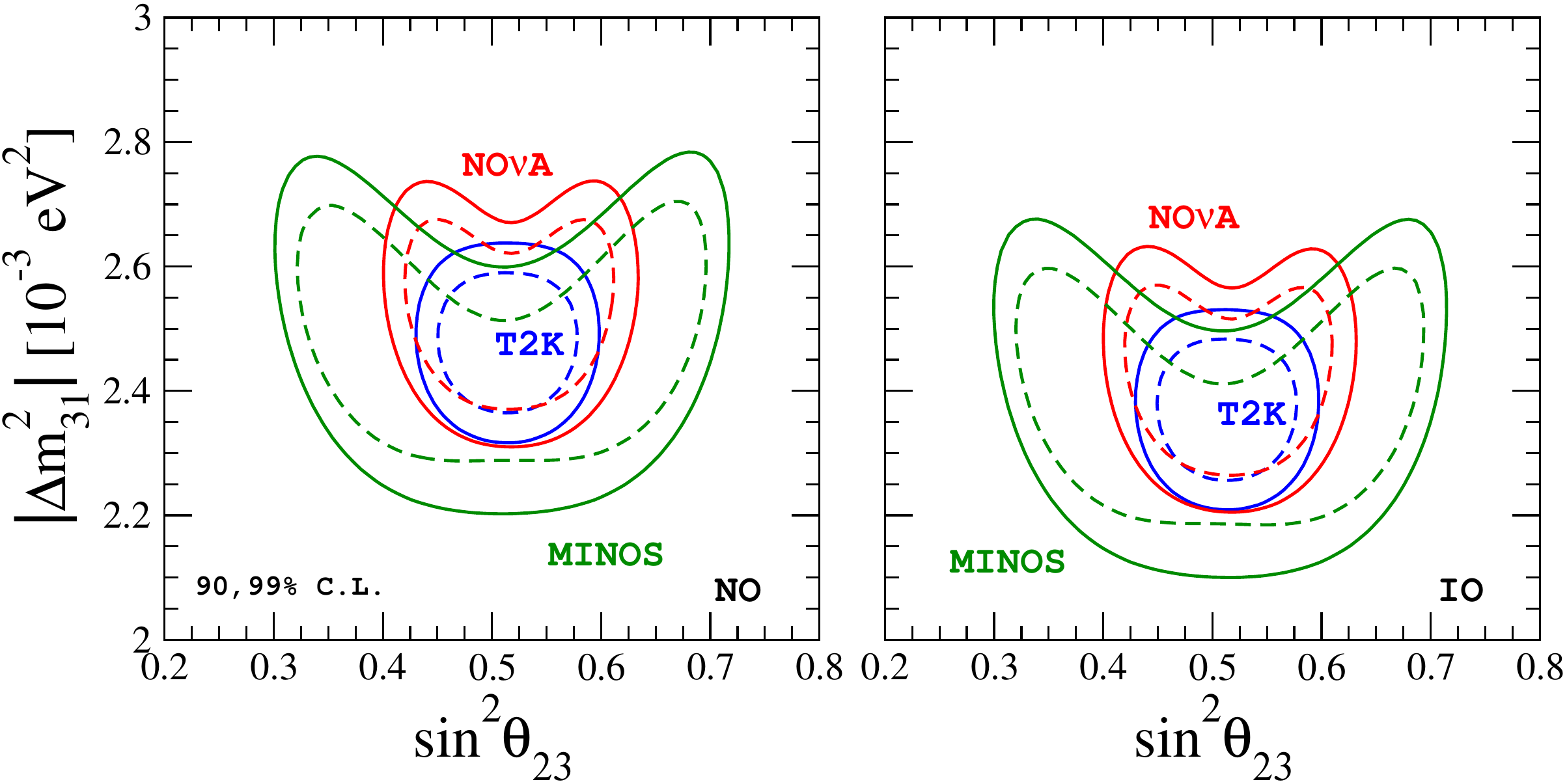}
   \caption{90 and 99\% C.L. (2 d.o.f.) allowed regions at the
     $\sin^2\theta_{23}$-- $\Delta m^2_{31}$ plane for normal (left)
     and inverted mass ordering (right) as restricted from the
     long--baseline experiments.}
   \label{fig:LBL}
 \end{figure}

\section{Global fit results}
\label{sec:global-fit-2017}

 We now describe the global results of our updated neutrino
  oscillation fit.  There are no significant changes derived from the
  new solar neutrino data, hence we move directly to the results for
  atmospheric neutrinos.
  Here there are new data from the ANTARES and IceCube collaborations
as well as from Super-Kamiokande phase IV.
  As seen in Fig.~\ref{fig:all-atm}, the 863-day atmospheric data from
  ANTARES and the 3-year data from IceCube DeepCore are enough to
  provide a determination of the atmospheric oscillation parameters.
 Note, however, that the determination of $\theta_{23}$
    from atmospheric data is still dominated by the analysis of
    Super-Kamiokande. 
  In any case, the neutrino telescope results are in complete
  consistency with what follows from the Super-Kamiokande atmospheric
  data, leading to a clear global picture for the all-atmospheric data
  fit, shown in Fig.~\ref{fig:all-atm}.

  Concerning the long--baseline accelerator data,
  Fig.~\ref{fig:LBL} shows the allowed regions by the latest NO$\nu$A
  and T2K neutrino results, as well as the older MINOS data sample. In
  comparison with Fig.~\ref{fig:all-atm}, one sees that atmospheric
  parameters are mainly constrained by long--baseline data, and that
  now all the results are in agreement with maximal atmospheric
  mixing.
  On the other hand, Fig.~\ref{fig:reactor} shows how the new reactor
  data, clearly dominated by Daya Bay, provide a significantly
  improved determination of $\theta_{13}$.
  It also illustrates the important role of reactor neutrino data in
  mapping out the allowed region of the atmospheric squared mass splitting
  parameter.\\[-.2cm]

  In what follows we highlight the main features of our neutrino
  oscillation global fit results, focusing upon the main open
  challenges of the three-neutrino picture: CP violation, the neutrino mass
  ordering and the $\theta_{23}$ octant problem.

\subsection*{Sensitivity to  CP violation}
\label{sec:delt}

\begin{figure}[t!]
 \centering
        \includegraphics[width=0.8\textwidth]{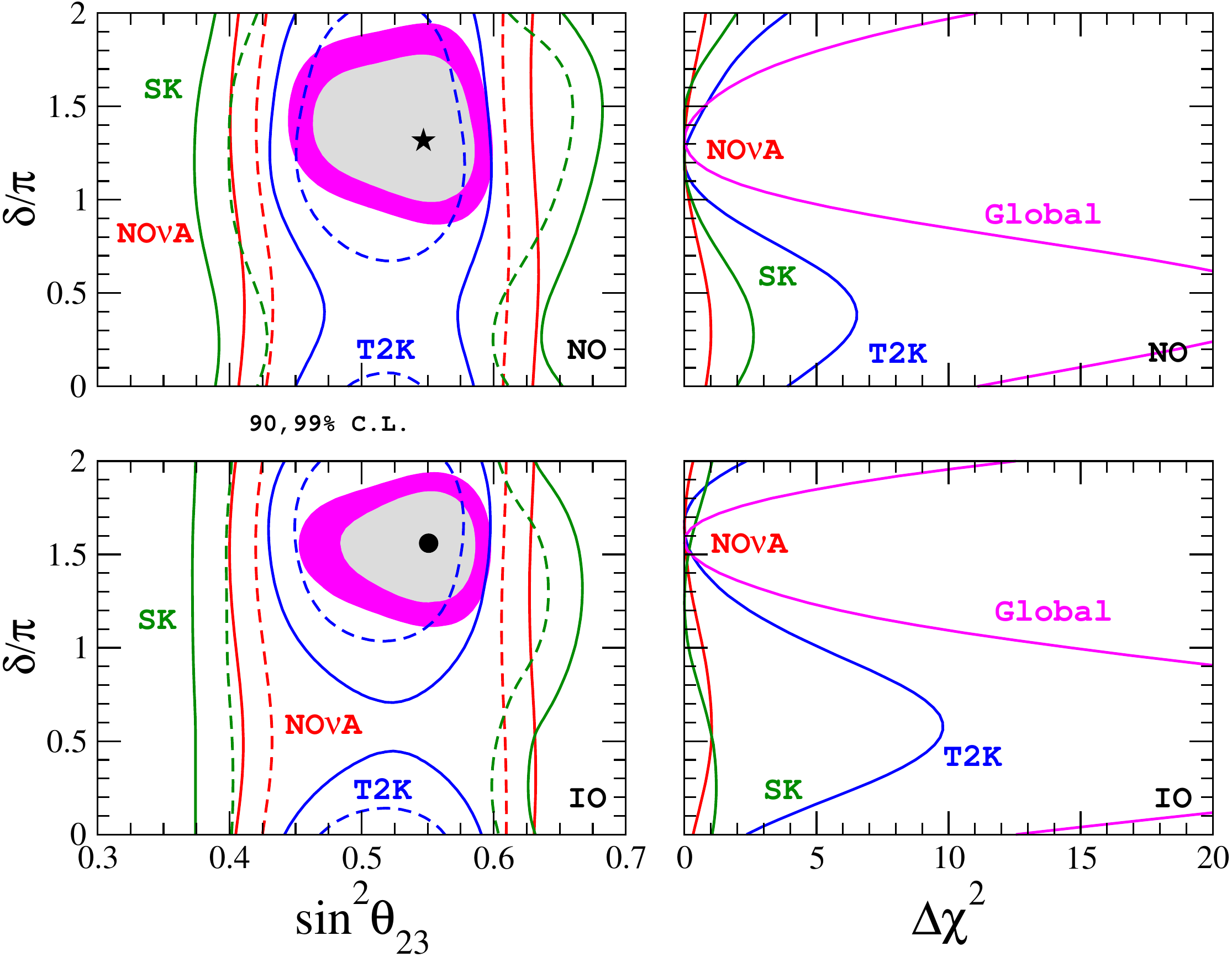}
        \caption{ Left: 90 and 99\% C.L. (2 d.o.f.) regions from
          T2K (blue lines) and NO$\nu$A (red) data, from the
          atmospheric Super-K results (green) and from the global fit
          of all the oscillation experiments (coloured regions). The
          star indicates the best fit point from our global analysis,
          found for normal mass ordering, while the black dot
          indicates the local minimum for inverted mass ordering.
          Right: $\Delta\chi^2$--profile as a function of the CP phase
          $\delta$ from T2K, NO$\nu$A and Super-K atmospheric (with
          the same color code as in the left panel) and from the global fit
          (magenta).  In both cases, the upper (lower) panels
          correspond to normal (inverted) mass ordering.}
        \label{fig:del-T2K-NOvA}
\end{figure}

Long--baseline neutrino oscillation data play an important role in
determining the CP violating phase, $\delta$.
In order to highlight this point we present the $\Delta\chi^2$--profile
for the CP phase, as determined from T2K, NO$\nu$A and Super-K
atmospheric data alone, as well as by the global oscillation data
sample, as shown in the right panels in Fig.~\ref{fig:del-T2K-NOvA}.
Note that here the $\Delta\chi^2$--profile has been obtained from the
local minimum for each mass ordering. 

This result shows how the current global sensitivity to the CP phase
is dominated by the T2K experiment, with  added rejection against
$\delta = \pi/2$ obtained after combining with the other experiments.
Indeed, we find that the combination with reactor data is crucial to
enhance the rejection against $\delta = \pi/2$.
As a result, we find that in the global analysis, $\delta = \pi/2$ is
disfavoured with $\Delta\chi^2 = 22.9$ (4.8$\sigma$) for normal
ordering.  The rejection against $\delta = \pi/2$ is found to be
stronger for inverted mass spectrum, where it is excluded with
$\Delta\chi^2 = 37.3$ (6.1$\sigma$), with respect to the minimum for
this ordering.  
As can also be seen from the figure, the current
preferred value of $\delta$ depends on the mass ordering, lying closer
to $3\pi/2$ for inverted ordering.  The current best fit values for
the CP violating phase are located at $\delta = 1.21\pi$ for NO and at
$\delta = 1.56\pi$ for IO.\\[-.2cm]

\subsection*{Neutrino mass ordering}
\label{sec:mass-order}

Concerning the sensitivity to the neutrino mass ordering, our global
fit shows for the first time a hint in favour of normal neutrino 
mass ordering, with inverted ordering disfavoured with
$\Delta\chi^2  = 11.7$ (3.4$\sigma$).  
In order to disentangle the origin of the preference for NO in our
global analysis, we display in Figs.~\ref{fig:sq23.sq13.del} and
\ref{fig:sq23.sq13} the allowed regions for $\theta_{23}$,
$\theta_{13}$ and $\delta$ for NO and IO for different data set
combinations: long--baseline data only, long--baseline plus
atmospheric, long--baseline plus reactors and the combination of all
data sets.  Down--triangles indicate the best fit points obtained in
the analysis of long--baseline data, squares correspond to the best
fit points derived from the combination of long--baseline plus
atmospheric, while up--triangles are the best fit point for
long--baseline plus reactor data.  The star and black dot follow the
same convention as in Fig.~\ref{fig:del-T2K-NOvA}.

\begin{figure}[t]
 \center
        \includegraphics[width=0.75\textwidth]{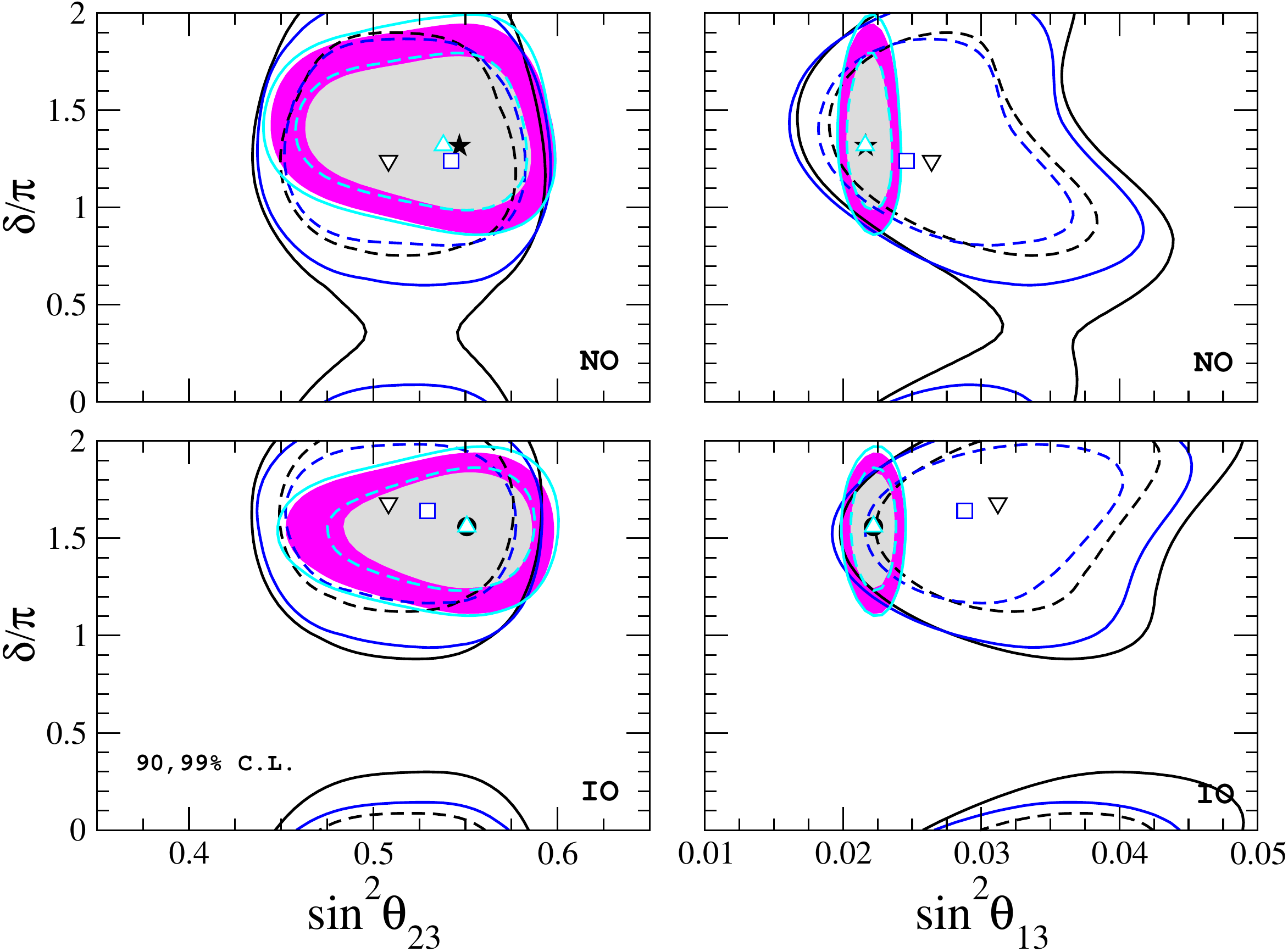}
        \caption{90 and 99\% C.L. (2 d.o.f.) allowed regions
          in the $\sin^2\theta_{23}$--$\delta$ (left) and
          $\sin^2\theta_{13}$--$\delta$ (right) planes from
          long--baseline data only (black lines), long--baseline plus
          atmospheric (blue), long--baseline plus reactors (cyan) and
          from the global fit of all experiments (coloured
          regions). Upper (lower) panels correspond to normal
          (inverted) mass ordering. The best fit points are indicated
          by black down--triangles (long--baseline data), blue squares
          (long--baseline plus atmospheric), cyan up--triangles
          (long--baseline plus reactors), as well as stars and black
          dots, following the same convention as in
          Fig.~\ref{fig:del-T2K-NOvA}. }
	\label{fig:sq23.sq13.del}
        \end{figure}

 The black lines in these figures delimit the allowed regions from the
 combination of all long--baseline data discussed above. 
 In principle, given the small impact of matter effects in
 the neutrino propagation at such baselines, one would
 expect a limited sensitivity of the current long--baseline
 experiments to the neutrino mass ordering. Indeed, this is confirmed
 by our independent analysis of T2K and NO$\nu$A data, that show only
 a slight preference for normal mass ordering at the level of
 $\Delta\chi^2$ $\sim$ 1. 
 However, the combined analysis of long--baseline and reactor data (see
 cyan lines in the figures) results in an enhanced sensitivity to the
 mass ordering which, in all cases, favours normal over inverted mass
 ordering. This happens due to the mismatch between the values of
 $\theta_{13}$ preferred by reactor and long--baseline experiments,
 which is larger for the inverted mass ordering, as shown in
 Figs.~\ref{fig:sq23.sq13.del} and \ref{fig:sq23.sq13}. 
 While in normal ordering the best fit for long--baseline experiments
 alone, $\sin^2\theta_{13}$ = 0.026, is relatively close to the global
 one, $\sin^2\theta_{13} \simeq$ 0.022, mainly constrained by
 reactors, this is not the case for inverted ordering, where
 long--baseline data prefer $\sin^2\theta_{13}$ = 0.031. 
 As a result, the combined analysis of reactor and long--baseline data
 shows better agreement under the normal mass ordering hypothesis.
 For instance, a combined analysis of the latest NO$\nu$A results with
 reactor data indicates a preference for normal ordering with
 $\Delta\chi^2 = 3.7$. 
 In the case of T2K, the combination with reactor data results in a
 stronger preference for normal over inverted mass ordering, with
 $\Delta\chi^2 = 5.3$. 
 This enhanced sensitivity to the mass ordering is due to the tension
 that exists between the value of the atmospheric mass splitting
 preferred by reactor, mainly Daya Bay, and T2K.  One finds that Daya Bay prefers a
 higher value for $\Delta m^2_{31}$ with respect to the one indicated
 by T2K, and the difference is larger for inverted mass ordering.
 The combined analysis of all long--baseline and reactor data yields a
 preference for normal mass ordering with $\Delta\chi^2 =
 7.5$. 

 By combining these data samples with atmospheric data, one gets the
 final global results indicated by the coloured regions in
 Figs.~\ref{fig:sq23.sq13.del} and \ref{fig:sq23.sq13}. 
 In principle, one may expect the largest sensitivity to the neutrino
 mass ordering to come from the observation of matter effects in the
 atmospheric neutrino flux. 
 However, we find that the neutrino telescope experiments IceCube
 DeepCore and ANTARES are not yet very sensitive to the mass
 ordering. In fact, the difference between normal and inverted mass
 ordering from the combined analysis of DeepCore and ANTARES is only
 $\Delta\chi^2 = 0.4$, obtained mainly from IceCube DeepCore data. 
On the other hand, the most recent analysis of atmospheric data sample of
 the Super-K experiment shows an enhanced sensitivity to the mass
 ordering compared to previous ones. 
 Indeed, Super-K data alone disfavours the inverted mass ordering with
 $\Delta\chi^2 = 3.5$. If a prior on the reactor mixing angle is
 imposed in the atmospheric data analysis, the sensitivity rises up to
 $\Delta\chi^2 = 4.3$~\cite{Abe:2017aap}.
 The effect of adding the atmospheric data to the global analysis is
 not very noticeable in Figs.~\ref{fig:sq23.sq13.del} and
 \ref{fig:sq23.sq13}, where the allowed regions with and without
 atmospheric data are similar.  However, the impact of atmospheric
 data in the global sensitivity to the mass ordering allows one to
 disfavour inverted mass ordering at $\Delta\chi^2 = 11.7$.
 This result is very relevant, since from the combination of the different
 types of neutrino experiments we can obtain for the first time a preference 
 for normal neutrino mass ordering slightly above 3$\sigma$.

\subsection*{The $\theta_{23}$ octant problem}
\label{sec:thet-octant-probl}

The role of long--baseline, atmospheric and reactor experiments in
selecting the $\theta_{23}$-octant is illustrated in
Figs.~\ref{fig:sq23.sq13.del} and \ref{fig:sq23.sq13}.  They stress
the complementary role of the different oscillation data samples on
the possible discrimination of the $\theta_{23}$ octant. In fact, as
noticed in Ref.~\cite{Huber:2009cw} and recently
in~Ref.~\cite{Chatterjee:2017irl}, an improved measurement of the
reactor angle helps resolving the atmospheric octant. 
From the figures, we see that the analysis of long--baseline data only
(indicated by black lines) shows a preference for values of
$\theta_{23}$ close to maximal mixing for the two mass orderings, with
the best fit points indicated by a down--triangle. For both mass orderings
we find a preferred value of $\sin^2\theta_{23} = 0.508$.  
The combination with the atmospheric data sets (illustrated by the
blue lines in the figures) provides a further constraint on the
allowed region for $\theta_{23}$.
Moreover, the inclusion of atmospheric data in the analysis produces a
shift of the best fit value of $\theta_{23}$ to larger values for
both mass orderings ($\sin^2\theta_{23} = 0.54$ for NO and
$\sin^2\theta_{23} = 0.53$ for IO), although values of $\theta_{23}$
in the first octant are still allowed with
 $\Delta\chi^2 \geq 0.8 \, (2.0)$ for NO (IO). 
 The combination with reactor experiments in the global neutrino fit
 (coloured regions in the figures) moves the best fit value of
 $\theta_{23}$ to larger values for both mass orderings, leading to
 $\sin^2\theta_{23} \simeq 0.55$ as  the preferred value.
 Moreover, reactor data also modify the preferred values of
 $\theta_{13}$ and $\delta$, from the values fixed by the combination
 of long--baseline and atmospheric data, as shown in
 Figs.~\ref{fig:sq23.sq13.del} and \ref{fig:sq23.sq13}.  
For both mass orderings $\delta$ is pushed towards $3\pi/2$, while the
reactor mixing angle is slightly shifted towards smaller values.
One should stress that reactor data, specially Daya Bay and RENO, are
crucial to the determination of the allowed region for $\theta_{13}$. 
Going back to the octant preference, we would like to remark that the
indications described above are still far from robust. Indeed, values
of the atmospheric mixing angle below $45\degree$ are allowed at
$\Delta\chi^2 \geq 1.6$ for the case of normal ordering and
  at $\Delta\chi^2 \geq 3.2$ for inverted ordering with respect
    to the minimum in this mass spectrum.
 
\begin{figure}[t]
 \centering
        \includegraphics[width=0.8\textwidth]{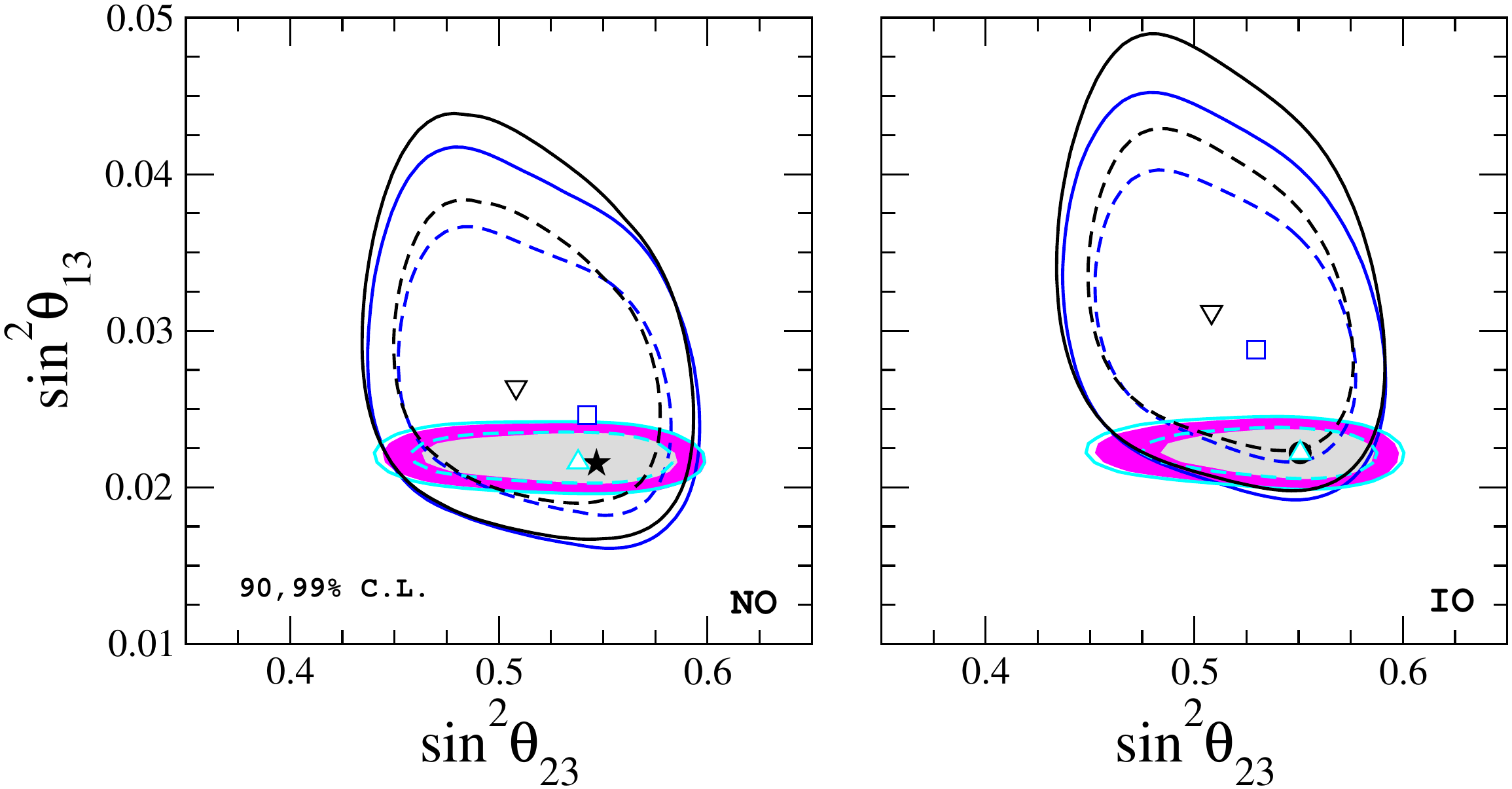}
        \caption{90 and 99\% C.L. (2 d.o.f.) regions from
          the combination of different neutrino data samples.  The
          convention used to indicate the regions and best fit points is the same
          as in Fig.~\ref{fig:sq23.sq13.del}.}
          \label{fig:sq23.sq13} 
          \end{figure}

Despite the recent progress on this matter, the octant discrimination
problem lies far beyond the current generation of neutrino oscillation
experiments, and will be a particularly stubborn problem in the years
to come.
On the positive side, however, it has been noted that the task of
octant discrimination and probing for leptonic CP violation in current
and future long--baseline experiments can be facilitated by prior
model-specific theoretical knowledge of the predicted pattern of
leptonic mixing.  See, as an example, Figure 1 given
in~\cite{Chatterjee:2017xkb} and the associated discussion.

\section{Summary and discussion}
\label{sec:summary-global-fit}

\begin{table}[t!]\centering
  \catcode`?=\active \def?{\hphantom{0}}
   \begin{tabular}{lccc}
    \hline
    parameter & best fit $\pm$ $1\sigma$ &   \hphantom{x} 2$\sigma$ range  \hphantom{x} &  \hphantom{x} 3$\sigma$ range  \hphantom{x}
    \\
    \hline
    $\Delta m^2_{21}\: [10^{-5}\eVq]$
    & 7.55$^{+0.20}_{-0.16}$  & 7.20--7.94 & 7.05--8.14 \\[3mm]  
    $|\Delta m^2_{31}|\: [10^{-3}\eVq]$ (NO)
    &  2.50$\pm$0.03 &  2.44--2.57 &  2.41--2.60\\
     $|\Delta m^2_{31}|\: [10^{-3}\eVq]$ (IO)
    &  2.42$^{+0.03}_{-0.04}$ &  2.34--2.47 &  2.31-2.51 \\[3mm] 
    $\sin^2\theta_{12} / 10^{-1}$
    & 3.20$^{+0.20}_{-0.16}$ & 2.89--3.59 & 2.73--3.79\\ 
    $\theta_{12} /\degree$ & 34.5$^{+1.2}_{-1.0}$ & 32.5--36.8 & 31.5--38.0 \\[3mm]  
     $\sin^2\theta_{23} / 10^{-1}$ (NO)
              &	5.47$^{+0.20}_{-0.30}$ & 4.67--5.83 & 4.45--5.99 \\
     $\theta_{23} /\degree$ &	47.7$^{+1.2}_{-1.7}$ &  43.1--49.8& 41.8--50.7\\ 
     $\sin^2\theta_{23} / 10^{-1}$ (IO)
              & 5.51$^{+0.18}_{-0.30}$ & 4.91--5.84 & 4.53--5.98\\
     $\theta_{23} /\degree$ & 47.9$^{+1.0}_{-1.7}$ & 44.5--48.9 & 42.3--50.7\\[3mm]  
    $\sin^2\theta_{13} / 10^{-2}$ (NO)
    & 2.160$^{+0.083}_{-0.069}$ &  2.03--2.34 & 1.96--2.41 \\
    $\theta_{13} /\degree$    & 8.45$^{+0.16}_{-0.14}$ & 8.2--8.8 & 8.0--8.9 \\ 	
    $\sin^2\theta_{13} / 10^{-2}$ (IO)
    & 2.220$^{+0.074}_{-0.076}$ & 2.07--2.36 & 1.99--2.44 \\ 
    $\theta_{13} /\degree$ & 8.53$^{+0.14}_{-0.15}$ & 8.3--8.8 &  8.1--9.0 \\[3mm] 
   $\delta/\pi$ (NO)
   	& 1.21$^{+0.21}_{-0.15}$ & 1.01--1.75 & 0.87--1.94 \\
   $\delta/\degree$	& 218$^{+38}_{-27}$ & 182--315 & 157--349\\  
    $\delta/\pi$ (IO)	
   	& 1.56$^{+0.13}_{-0.15}$ & 1.27--1.82 & 1.12--1.94 \\
   $\delta/\degree$	& 281$^{+23}_{-27}$ & 229--328 & 202--349\\	
       \hline
     \end{tabular}
     \caption{ \label{tab:sum-2017} 
        Neutrino oscillation parameters summary determined from this 
        global analysis. The ranges for inverted ordering refer to the 
        local minimum for this neutrino mass ordering.
     }
\end{table}

We have discussed in detail the origin of the mass ordering, CP
violation and octant discrimination, analyzing the interplay among the
different neutrino oscillation data samples.
The results obtained in our global fit are summarized in Table
\ref{tab:sum-2017} as well as Figs.~\ref{fig:panel-dchi2} and
\ref{fig:2dim} for normal and inverted mass ordering.
Some comments are in order. 

First we note that the improved precision on $\theta_{13}$ follows
mainly from the Daya Bay and RENO data.
Thanks to the combination of T2K neutrino and antineutrino data, we
have now an improved sensitivity to CP violation. Indeed, T2K is the
first experiment showing a sensitivity on its own, excluding some
values of $\delta$ before combining with reactor data. In this
analysis, we have obtained a strong preference for values of the CP
phase in the range $[\pi, 2\pi]$, excluding values close to $\pi/2$ at
more than 4$\sigma$.
Concerning the octant of $\theta_{23}$, this global analysis prefers
the second octant slightly, in agreement with the previous one in
Ref.~\cite{Forero:2014bxa}.
We have found that for normal neutrino mass ordering the upper
atmospheric octant is now preferred with $\Delta\chi^2 = 1.6$, while
for the case of inverted ordering, values of the atmospheric mixing
angle in the lower octant are allowed with $\Delta\chi^2 \geq 3.2$.
More remarkably, our global analysis favours for the first time the
normal mass ordering over the inverted one at 3.4$\sigma$.  As
discussed in the previous section, part of the sensitivity to the mass
ordering comes from the more recent atmospheric analysis of Super-K.
This new analysis shows a preference for normal over inverse mass
ordering with $\Delta\chi^2 = 3.5$.
On the other hand, a mismatch between the values of $\theta_{13}$
  preferred by long--baseline and reactor data (larger for IO) also gives a
  relevant contribution to the global sensitivity to the mass
  ordering. This effect is also enhanced due to a tension between the preferred values of the atmospheric mass splitting by T2K and reactor experiments.
  
\begin{figure}[H]
 \centering
\includegraphics[width=0.9\textwidth]{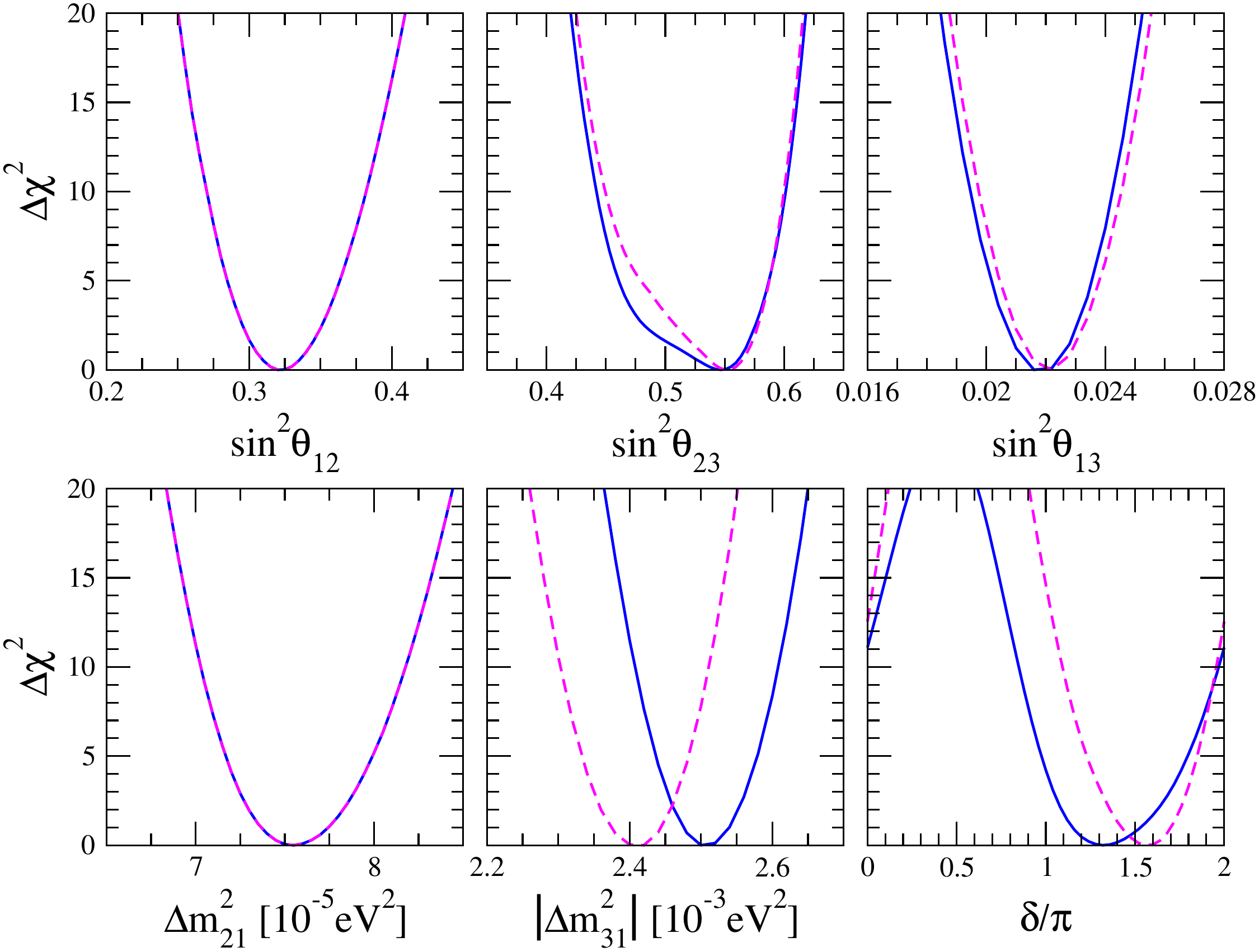}
\caption{\label{fig:panel-dchi2} Summary of neutrino oscillation
  parameters, 2018. Blue (solid) lines correspond to NO and magenta
  (dashed) lines to IO.  Notice that the $\Delta\chi^2$--profiles
    for inverted ordering are plotted with respect to the minimum for
    this neutrino mass ordering.}
\end{figure}

\black In short, we have seen how the precision in the determination
of the best-known oscillation parameters has improved thanks to the
recent long--baseline neutrino oscillation and reactor data.  Also the
sensitivity to mass ordering, CP violation and the octant of the
atmospheric angle has improved, although we are still quite far from a
robust measurement, especially of the octant.  The presence
of new physics beyond the Standard Model may affect significantly the
results obtained within the current neutrino oscillation picture.  For
example, nonstandard neutrino interactions with matter and non-unitary
neutrino mixing, expected in seesaw models of neutrino mass
generation, may significantly reduce the sensitivities. Conversely,
however, such well--motivated beyond--standard scenarios can also bring
in new opportunities for current and future long--baseline neutrino
oscillation experiments.
\begin{figure}[H]
 \centering
    \includegraphics[width=0.6\textwidth]{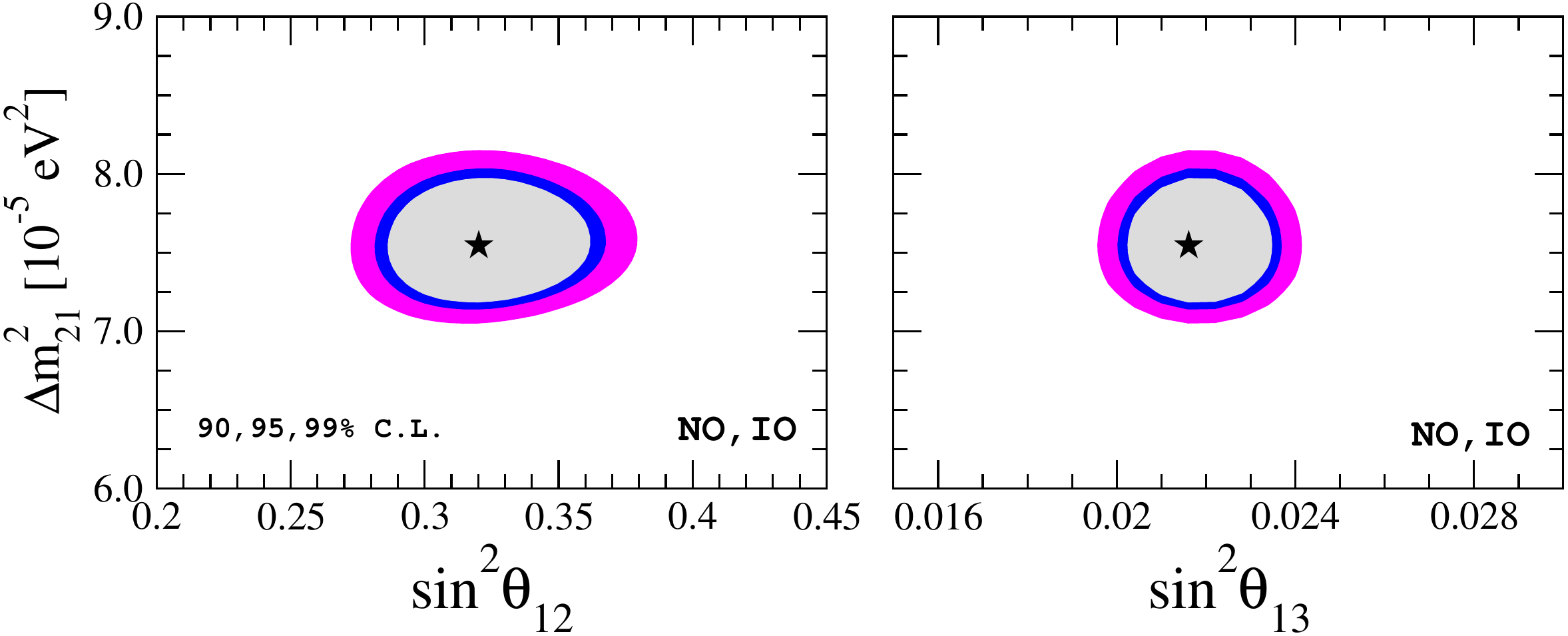}
    \includegraphics[width=0.6\textwidth]{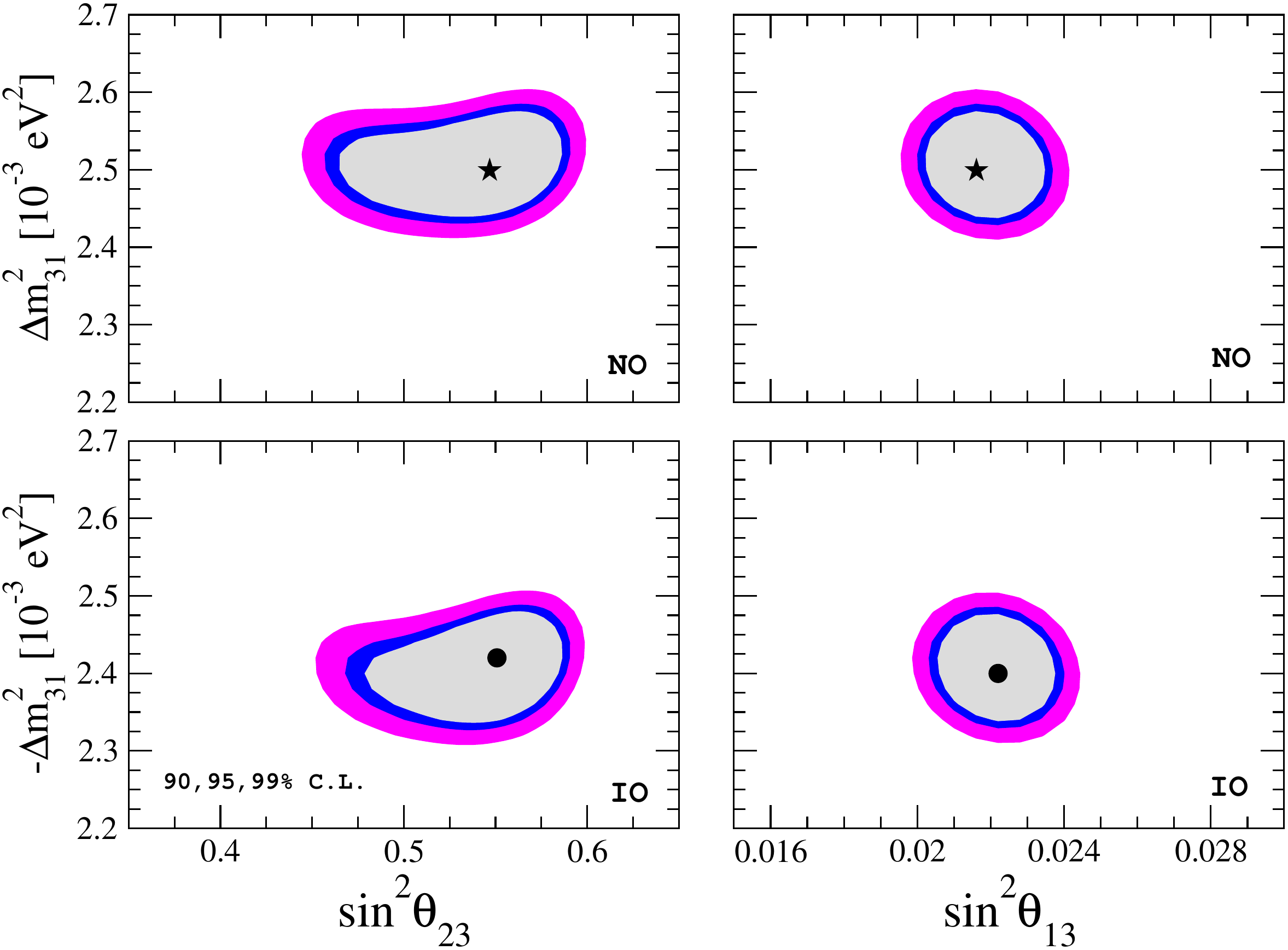}
    \includegraphics[width=0.6\textwidth]{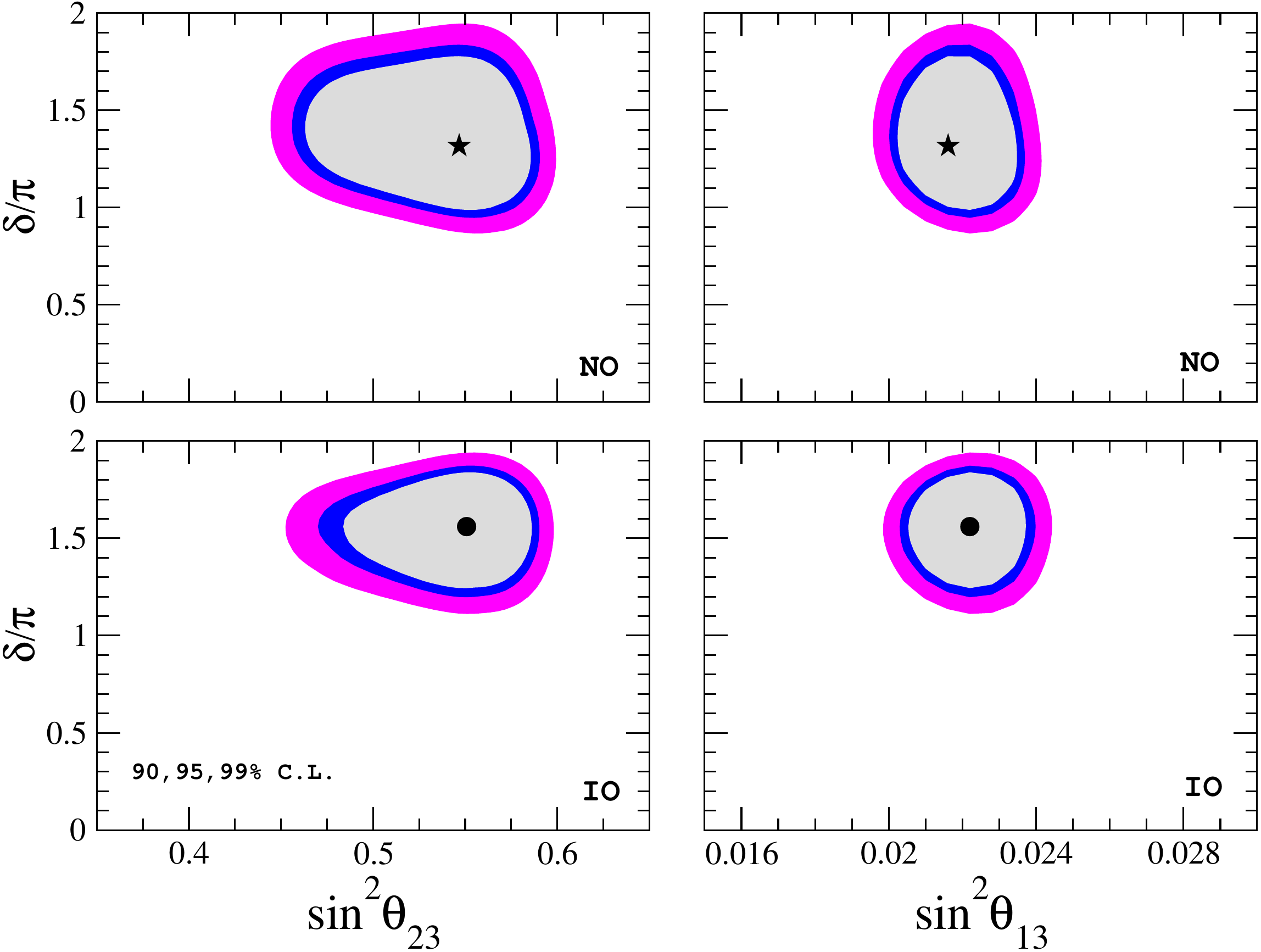}
    \caption{\label{fig:2dim} Global fit summary 2018.  In the two
      four-panel figures, the upper ones correspond to normal ordering
      and the lower ones to inverted mass ordering. Global fit regions
      correspond to 90, 95 and 99\% C.L. (2 d.o.f.). As in
      Fig.~\ref{fig:panel-dchi2}, regions for inverted ordering 
      are plotted with respect to the minimum for
    this neutrino mass ordering.}
\end{figure}
}

\section*{Acknowledgments}

The authors would like to thank Jason Koskinen of the IceCube
collaboration, Juande Zornoza of the ANTARES collaboration and
Federico S\'anchez from T2K, as well as Jordi Salvado and Thomas
Schwetz for useful discussions.
Work supported by MINECO grants FPA2014-58183-P,
Multidark-CSD2009-00064, SEV-2014-0398, and the PROMETEOII/2014/084
and GV2016-142 grants from Generalitat Valenciana.
MT is also supported a Ram\'{o}n y Cajal contract (MINECO).
PFdS is supported by the Spanish grant FPU13/03729 (MECD).
DVF is thankful for the support of FAPESP Grants No. 2014/19164-6
and 2017/01749-6, and also to FAEPEX Grant No 2391/17 for partial
support. DVF was also supported by the U.S. Department Of Energy under
contracts DE-SC0013632 and DE-SC0009973.

\begingroup
\raggedright
\sloppy

\bibliographystyle{apsrev}

\end{document}